\documentclass[aps,prl,floats,twocolumn,floatfix,showpacs]{revtex4}
\usepackage{epsf,graphicx}
\usepackage{amssymb}
\usepackage{amsmath}
\usepackage{eepic}
\usepackage[dvips]{color}
\newlength{\piclen}
\begin{document}

\title{The Fano effect in the point contact spectroscopy of heavy electron materials}

\author{Yi-feng Yang}
\affiliation{Los Alamos National Laboratory, Los Alamos, NM 87545 and\\
Department of Physics, University of California, Davis, CA 95616}

\begin{abstract}
We show that Fano interference explains how point contact spectroscopy in heavy electron materials probes the emergence of the Kondo heavy electron liquid below the same characteristic temperature T* as that seen in many other experiments, and why the resulting measured conductance asymmetry reflects the universal Kondo liquid behavior seen in these. Its physical origin is the opening of a new channel for electron tunneling beyond that available from the background conduction electrons. We propose a simple phenomenological expression for the resulting Fano interference that provides a good fit to the experimental results for CeCoIn$_5$, CeRhIn$_5$ and YbAl$_3$, over the entire range of bias voltages, and deduce a life-time of the heavy quasiparticle excitations that agrees well with recent state-of-the-art numerical calculations.
\end{abstract}


\maketitle

Heavy electron materials have a rich phase diagram showing a competition between
antiferromagnetism and unconventional superconductivity and quantum critical
behavior. Although the underlying physics responsible for this competition is still unclear, it appears to be primarily associated with the heavy electron, or Kondo liquid, that emerges from the collective hybridization of light conduction
electrons with the local f-moments \cite{Nakatsuji2004,Curro2004,Yang2008,Yang2008b}. Understanding the nature and consequence of this hybridization is therefore a central task in the field of
heavy electron physics.

Point contact spectroscopy (PCS) \cite{Naidyuk2005} probes the low energy collective excitations (such as phonons) and may be expected to provide important information for our understanding of the low energy physics of heavy electron materials. For superconductors, PCS provides a quantitative measure of Andreev reflection and helps determine the superconducting order parameters and the paring mechanism \cite{Rourke2005,Park2005}. Especially, a universal asymmetry has been observed in the point contact tunneling experiments of high T$_C$ superconductors \cite{Kirk1987}, in contrast to what is expected for Bardeen-Cooper-Schrieffer type superconductors.

For decades, a similar conductance asymmetry has also been observed in many heavy electron materials such as CeCu$_{6}$ \cite{Moser1986} and URu$_2$Si$_2$ \cite{Hasselbach1992} but its origin has not been understood. For example, the fact that the asymmetry is practically independent of the material of the metallic tips suggests that it is intrinsic and excludes previous explanations based on self-heating effects \cite{Nowack1997}. On the other hand, the usual tunneling model requires an unrealistic background density of states (DOS) to explain the experimental data. In CeCoIn$_5$, the DOS derived from it has a broad maximum below the Fermi energy \cite{Shaginyan2007,Park2008}, in
contradiction with the theoretical expectations and numerical calculations that show a sharp quasiparticle peak developing at low temperatures well above the Fermi energy \cite{Shim2007}.

An essential clue to the underlying physics comes from two recent observations concerning the point contact spectroscopy of CeCoIn$_5$. First, Park {\em et al.} \cite{Park2008} measured, for the first time, the temperature variation of the conductance asymmetry in the normal state. It was later pointed out \cite{Curro} that this temperature variation was remarkably similar to that seen in the Knight shift and other experiments that measure the effective DOS of
the Kondo liquid that emerges below a characteristic temperature T* \cite{Yang2008}. This observation suggests an intimate connection between the asymmetry seen in PCS results and the emergent heavy fluid and establishes that the pronounced conductance asymmetry must be an intrinsic property of the heavy electron material. Second, for small bias voltages at a single low temperature, a simple Fano line-shape was shown to provide a fit to the experimental data \cite{Park2008}, although its origin was not specified. In the present communication, we argue that the measured asymmetry originates in the interference between the heavy and light (conduction) electron channels on
electron tunneling. We propose a simple phenomenological generalization of the standard Fano expression, and show that it provides an excellent fit to the experimental data for a broad range of voltages and temperatures below T*.

Heavy electron materials are usually modelled as a Kondo lattice of local f-moments coupled antiferromagnetically to conduction electrons. Although the basic physics differs from that of a single Kondo impurity, it exhibits similar low energy quasiparticle excitations that form a narrow f-electron band. In the slave boson formulation \cite{Millis1987}, the Kondo lattice model takes the form:
\begin{equation}
H=\sum_{k,m}\left[\epsilon_{k}c^\dagger_{km}c_{km}
+\epsilon_{0}f^\dagger_{km}f_{km}
+\tilde{V}(c^\dagger_{km}f_{km}+\text{H.c.})\right],
\label{Eq:Ham}
\end{equation}
where $c_{km}$ and $f_{km}$ correspond to the m-th fermionic operator of
the conduction electrons and the f-spins, respectively. $\epsilon_{0}$ and
$\tilde{V}$ are the renormalized f-level and the c-f coupling. This defines a
temperature T* above which the conduction and f-electrons are effectively decoupled and below which Eq.~(\ref{Eq:Ham}) can be diagonalized in terms of new fermionic operators \cite{Barzykin2006},
\begin{eqnarray}
d_{1km} &=& u_k f_{km} + v_k c_{km},\nonumber\\
d_{2km} &=& -v_k f_{km} + u_k c_{km},
\end{eqnarray}
with $u_k^2=[1+(\epsilon_{k}-\epsilon_{0})/E_k]/2$, 
$v_k^2 =[1-(\epsilon_{k}-\epsilon_{0})/E_k]/2$, 
and $E_k=[(\epsilon_{k}-\epsilon_{0})^2+4\tilde{V}^2]^{1/2}$. The new fermionic
operators describe two noninteracting hybridization bands with the energies $\epsilon_{1k}=(\epsilon_k+\epsilon_0-E_k)/2$ and
$\epsilon_{2k}=(\epsilon_k+\epsilon_0+E_k)/2$. In point contact experiment, a metallic tip is added to the system with a transfer Hamiltonian
\begin{equation}
H_t = \sum_{km}\left(M_{fkm}f_{km}^\dagger t + M_{ckm}c_{km}^\dagger t + \text{H.c.}\right),
\end{equation}
where $t$ is the fermionic operator for the tunneling state of the tip. The tunneling matrix elements to the hybridization bands are hence given by
\begin{eqnarray*}
\lefteqn{|(d_{1km}|H_t|t)|^2 =|u_k(f_{km}|H_t|t)+v_k(c_{km}|H_t|t)|^2}\\
&&=\left|\,q+\frac{v_k}{u_k}\right|^2|u_k|^2|M_{ckm}|^2 =\frac{|q-\tilde{E}_{1k}|^2}{1+\tilde{E}_{1k}^2}|M_{ckm}|^2,\\
\lefteqn{|(d_{2km}|H_t|t)|^2 = |-v_k(f_{km}|H_t|t)+u_k(c_{km}|H_t|t)|^2}\\
&&= \left|\,q-\frac{u_k}{v_k}\right|^2|v_k|^2|M_{ckm}|^2=
\frac{|q-\tilde{E}_{2k}|^2}{1+\tilde{E}_{2k}^2}|M_{ckm}|^2,
\end{eqnarray*}
where $\tilde{E}_{ik}=(\epsilon_{ik}-\epsilon_0)/\tilde{V}$ and the Fano parameter $q=M_{fkm}/M_{ckm}$ is the ratio of the tunneling couplings to the itinerant f- and conduction electrons. Following Ref.~\cite{Harrison1961} and using Fermi's golden rule, we get the total differential conductance
\begin{eqnarray}
G(V,T)&=&g_0+\int\,g_I(E) T(E)\frac{df(E-V)}{dV}dE \nonumber\\
&\approx&g_0+g_I T(V),
\end{eqnarray}
with
\begin{equation}
T(E)=\frac{|q-\tilde{E}|^2}{1+\tilde{E}^2},
\label{Eq:TE}
\end{equation}
that has a simple Fano line-shape \cite{Fano1961} with a normalized energy
$\tilde{E}=(E-\epsilon_0)/\tilde{V}$. Here $f(E)$ is the Fermi distribution function and $g_0$ denotes a constant background conductance. $g_I(E) \propto \rho_{t}\sum_{ikm}|M_{ckm}|^2\delta(E-\epsilon_{ik})$ has typically a complicated form depending on the density of states $\rho_{t}$ of the metallic tip, the band structure of the system, and details of the tunneling barrier \cite{Harrison1961} and is assumed to be a constant in the following for simplicity. $\epsilon_0$ is the position of the Kondo liquid resonance which must be above the Fermi energy for Ce-compounds in analogy to the numerical results for CeIrIn$_5$ \cite{Shim2007}. Its exact value may be approximated as a constant between 0 and T*. For CeCoIn$_5$, this means 0$<\epsilon_0<4\,$meV.

The above formula is similar to that derived for a single impurity Kondo system \cite{Madhavan2001}. For a single Kondo impurity on a metallic surface, a Fano interference for tunneling into the local Kondo resonance has been measured in STM experiments \cite{Nagaoka2002}. Fano interference has also been observed in
a Kondo quantum dot embedded into one arm of an Aharonov-Bohm ring \cite{Kobayashi2003}. Our derivation shows that a similar Fano effect is expected below a characteristic temperature T* in the point contact spectroscopy of a Kondo lattice system due to the c-f hybridization. In reality, heavy electron materials invlove other complicated effects such as multi Fermi surface, anisotropic hybridization and antiferromagnetic exchange correlations among f-electrons that are beyond the simple mean field Hamiltonian in Eq.~(\ref{Eq:Ham}). The strong electronic correlations may introduce decoherence of tunneling electrons and destroy the Fano interference. This leads to a complex $q=q_1+iq_2$ beyond the simple ratio $M_{fkm}/M_{ckm}$. The imaginary part $q_2$ contributes a Lorentzian term $q_2^{\,2}/(1+\tilde{E}^2)$ to the total conductance, representing direct tunneling into the heavy electron states around $\epsilon_0$ without interference. Moreover, a simple Fano line-shape does not display asymmetry at large bias limit, $G(V)=G(-V)$ at $|V|\gg |\epsilon_0|, \tilde{V}$, while the experimentally observed conductance asymmetry extends over the whole spectra \cite{Park2008}. To account for all these effects beyond the simple hybridization picture described above, we redefine $\tilde{E}=(E-\epsilon_0)/\Gamma(V,T)$ in Eq.~(\ref{Eq:TE}) and introduce a phenomenological parameter $\Gamma(V,T)$ in such a way that $\Gamma(V,T)$ is asymptotically proportional to $|V|$ at large bias. The simplest way to achieve this is to take
\begin{equation}
\Gamma(V,T)=\sqrt{(aV)^2+\gamma^2},
\label{Eq:Gamma}
\end{equation}
with $a\sim 0.5$ and $\gamma(T)$ denoting the zero bias scattering rate within the Kondo liquid channel.

Physically, one expects that as energy and temperature increase, electrons injected into the heavy Kondo liquid will be more strongly scattered due to the strong electronic correlations. Excited crystal field states and voltage dependent tunneling matrix may also be the origins of Eq.~(\ref{Eq:Gamma}).
Still another possible explanation for the above bias dependent $\Gamma$ is the energy relaxation of the nonequilibrium electrons that leads to an effective broadening of the local heavy quasiparticle spectra. In the diffuse regime, injected electrons are strongly scattered close to the point contact, resulting in local heating effects with an effective temperature $T_{pc}$ that is related to the bias voltage by $T_{pc}^2=T^2+V^2/4L$ where $L$ is the Lorenz number of the material, which can be approximated by its Sommerfeld value $L_0=\pi^2/3$ for non-interacting electrons, For CeCoIn$_5$, we find an almost constant Lorenz number $L\approx L_0$ in the whole temperature range \cite{Onose2007}. For CeRhIn$_5$, the Lorenz number also approaches $L_0$ above $8\,$K \cite{Paglione2005}. These examples suggest that it is physically reasonable to approximate $a$ by a constant $\sim0.5$. Although heating effects may be absent in the ballistic regime, the nonequilibrium electrons tunneling into the heavy electron states suffer strong electronic correlations and are strongly scattered. This may lead to a similar energy relaxation and result in a similar broadening of the Kondo liquid spectra as is shown in Eq.~(\ref{Eq:Gamma}) close to the point contact. In the following, we take $a=0.5$ for simplicity. A different value of $a$ may be possible based on an improved understanding of the underlying physics. Since a microscopic theory of heavy electron materials is not yet available, the above equations provide a phenomenological picture that may shed light on future investigations.

\begin{figure}[t]
{\includegraphics[width=7.4cm,angle=0]{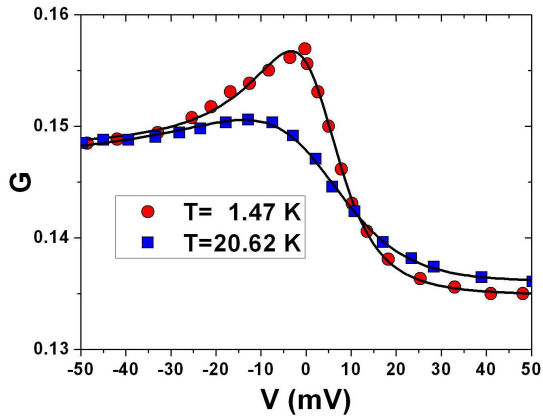}}
\caption{(Color online)
{Point contact spectra of CeCoIn$_5$ at $1.47\,$K and $20.62\,$K from -$50\,$mV to $50\,$mV \cite{Park2008}. A conductance asymmetry is clearly seen over the whole bias range up to $\pm$50 mV. The solid lines are the theoretical curves with $\epsilon_0= 3\,$meV, $a=0.5$ and $q_1=0.5$. Other parameters are $\gamma=12\,$meV, $g_I/g_0=0.14$, $q_2=1.26$ at $T=1.47\,$K and $\gamma=24\,$meV, $g_I/g_0=0.1$, $q_2=1.1$ at $T=20.62\,$K.}
\label{Fig:Fig1}}
\end{figure}

Besides $a\!\approx\!0.5$ and $\epsilon_0\!\sim\,$T*, three more constraints help to determine the other five parameters $g_0$, $g_I$, $q_1$, $q_2$ and $\gamma$. The average conductance at negative and positive large bias limit gives roughly an overall scale factor $g_0\approx (G^{+}+G^{-})/2$, while their difference $G^{-}-G^{+}\approx 4g_Iq_1 a/(1+a^2)$. By using $d\,T(V)/dV=0$, the peak position $V_{p}$ in the conductance spectrum gives the third constraint so that $(V_p-\epsilon_0)/\sqrt{\gamma^2+(aV_p)^2}$ is determined by $q_1$ and $q_2$. Hence only two free parameters $q_2$ and $\gamma$ (or $q_1$), among all seven parameters in our formula, are left to be determined by the fit. While $q_2$ leads to a small Lorentzian contribution, the resulting $\gamma$ must agree with numerical calculations. The validity of our formula for describing heavy electron point contact spectroscopy is therefore verified by its success in fitting to the experimental data at different temperatures for different materials, and the conductance asymmetry that is not seen in junctions of simple metals only shows up as a result of the characteristic heavy electron physics in Kondo lattice systems. In the following, we use Eqs.~(\ref{Eq:TE}) and (\ref{Eq:Gamma}) to study the point contact spectra of three different heavy electron materials, CeCoIn$_5$, CeRhIn$_5$ and YbAl$_3$.

\begin{figure}[t]
{\includegraphics[width=7.2cm,angle=0]{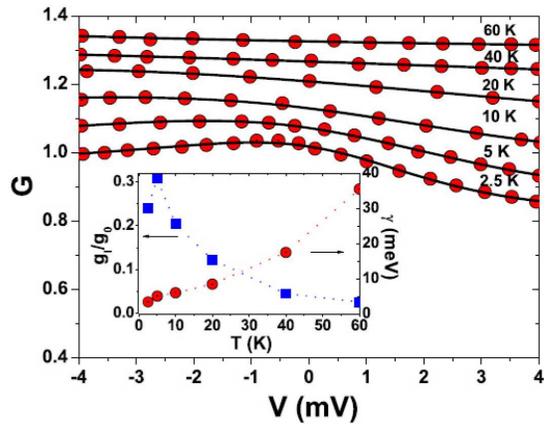}}
\caption{(Color online)
{Point contact spectra of CeCoIn$_5$ from -$4\,$mV to $4\,$mV at different temperatures \cite{Park2008}. A conductance asymmetry is developed below T*=$60\,$K. The solid lines are our fit with $\epsilon_0=0.8\,$meV and $a=0.5$. The best fit gives $q_1\!\sim\,$0.5 and $q_2\!\sim\,$1. The inset plots the temperature dependence of $\gamma$ and $g_I/g_0$. Since $q_1$ is roughly a constant, $g_I/g_0$ follows approximately the conductance asymmetry.}
\label{Fig:Fig2}}
\end{figure}

{\it CeCoIn$_5$.---}To show the necessity of introducing a bias dependent $\Gamma$, in Fig.~\ref{Fig:Fig1} we plot the conductance spectra of CeCoIn$_5$ from -$50\,$meV to $50\,$meV \cite{Park2008}. The differential conductance approaches different values at large positive and negative bias voltages. The asymmetry extends over the whole spectra and the typical Fano dip smears out. Thus there is a deviations from the simple Fano line-shape which has motivated us to put forward our modified formula. Taking for example $\epsilon_0=3\,$meV and $a=0.5$, we obtain good fits for both temperatures over the whole bias range
as plotted in Fig.~\ref{Fig:Fig1}. Below T$_c=2.3\,$K, Andreev reflection also has a small contribution within $\pm1\,$meV. Since we only focus on the normal phase, this small term is neglected for simplicity.

To study the temperature variation of the parameters, we fit the spectra for a variety of temperatures below T* in Fig.~\ref{Fig:Fig2}. The experimental data are only available between -$4\,$mV and $4\,$mV in the literature \cite{Park2008}. The conductance asymmetry decreases with increasing temperature and vanishes at $\sim60\,$K, in good agreement with the c-axis Kondo liquid temperature T* estimated from Knight shift anomaly \cite{Yang2008} and the coherence temperature in the magnetic resistivity \cite{Zapf2001}. Due to the small bias range of the data, the fit is sensitive to the value of $\epsilon_0$. The conductance may be more affected by the detail of the heavy quasiparticle band and slightly different values of $\gamma$ and $\epsilon_0$ may be required from those used in Fig.~\ref{Fig:Fig1}. But the overall behavior with temperature is the same.

The best fits give rise to a small Fano parameter $q_1\approx 0.5$, indicating that the metallic tip is more strongly coupled to the conduction electrons than the heavy electrons \cite{Madhavan2001}, a result expected from the large mismatch of the heavy electron velocities. The resonance width $\gamma$ at zero bias increases monotonically from $\sim5\,$meV at very low temperature to $\sim35\,$meV at $\sim60\,$K, implying an increasing broadening of the heavy quasiparticle excitations, similarly to that seen in the single impurity Kondo resonance. In the numerical calculations for CeIrIn$_5$, the width of the heavy quasiparticle spectra is found to be $\sim3\,$meV at very low temperature and increases to $\sim100\,$meV at $300\,$K \cite{Shim2007}. Given the similarity between CeCoIn$_5$ and CeIrIn$_5$, we are able to obtain quantitative agreement between our deduced value and that given by numerical calculations. Due to the small bias range and experimental errors, other different values of the fitting parameters may also lead to equally good fits. However, the agreement with theoretical and numerical expectations supports our proposed scenario.

\begin{figure}[t]
{\includegraphics[width=7.4cm,angle=0]{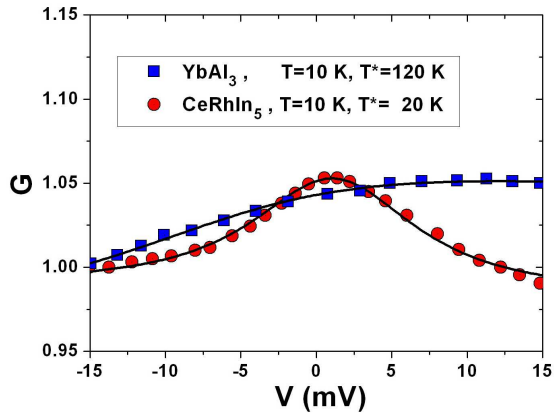}}
\caption{(Color online)
{Point contact spectra from -$15\,$mV to $15\,$mV at $10\,$K for both CeRhIn$_5$ and YbAl$_3$ \cite{Park2008b}. The solid lines are our fit with $a=0.5$. Other parameters are $\epsilon_0=1.5\,$meV, $\gamma=6.8\,$meV, $q_1=0.84$, $q_2=3.7$, and $g_I/g_0=0.0067$ for CeRhIn$_5$ and $\epsilon_0=-10\,$meV, $\gamma=21.9\,$meV, $q_1=-0.6$, $q_2=0.83$, and $g_I/g_0=0.06$ for YbAl$_3$.}
\label{Fig:Fig3}}
\end{figure}

{\it CeRhIn$_5$ and YbAl$_3$.---}Fig.~\ref{Fig:Fig3} shows the point contact spectra of CeRhIn$_5$ and YbAl$_3$ at $10\,$K \cite{Park2008b}. For CeRhIn$_5$, we take the quasiparticle energy $\epsilon_0=1.5\,$meV which is approximately T*=$20\,$K known from Hall anomaly and other experimental probes \cite{Yang2008}. The best fit with $a=0.5$ results in a zero-bias resonance
width $\gamma\sim6.8\,$meV, a reasonable value if we take into account the broadening at a temperature of half T*. The positive $\epsilon_0$ reflects a quasiparticle resonance above the Fermi energy in CeRhIn$_5$, similar to that in CeCoIn$_5$. 

On the other hand, YbAl$_3$ has a large T*$\sim$120 K \cite{Yang2008b} so that the Kondo liquid is well developed at $10\,$K. If the overall temperature dependence of the conductance asymmetry is known, it is expected to almost saturate at this temperature. Taking $a=0.5$ and $\epsilon_0=-10\,$meV, we find a large $\gamma=21.9\,$meV, consistent with the large T*. The negative resonance energy $\epsilon_0$ indicate the hole nature of the heavy quasiparticles in YbAl$_3$.

In our theory, the conductance asymmetry is intimately related to the heavy Kondo liquid that emerges at the characteristic temperature T*. Since T* can be probed in other ways such as a Knight shift anomaly and coherence seen in the optical conductivity \cite{Yang2008b}, we predict that the onset temperature
of the asymmetry must agree with that of the Knight shift anomaly, as well as the coherence temperature seen in the optical conductivity and the magnetic resistivity. Future point contact experiments will verify our prediction, that has already been found to apply to CeCoIn$_5$. We conclude that the point contact spectroscopy provides a quite useful way to determine T*. Since T* is the single characteristic temperature that governs the universal behavior
of the emergent heavy Kondo liquid, experiments that accurately determine this energy scale will definitely help us understand the physics of heavy electron materials.

In conclusion, we explain the point contact spectroscopy of heavy electron materials by a Fano interference effect of the tunneling electrons into the emergent hybridization bands. The conductance asymmetry is an essential result of the emergent heavy fluid and hence an intrinsic feature of heavy electron materials. Due to strong electronic correlations, a modified Fano line-shape is proposed and found to fit well all the experimental data for three different kinds of materials. The parameters obtained here are in good agreement with those given by numerical calculations. The point contact spectroscopy therefore provides important information on the low temperature physics of heavy electron materials. Our theory can be easily applied to other materials.

We wish to thank David Pines, Vladimir Sidorov, and Joe D Thompson for discussions. This research was supported by an ICAM Fellowship, UC Davis, and the Department of Energy.

\end{document}